\documentclass{article}%
\usepackage{amsmath}
\usepackage{amsfonts}
\usepackage{amssymb}
\usepackage{graphicx}%
\providecommand{\U}[1]{\protect\rule{.1in}{.1in}}
\newtheorem{theorem}{Theorem}

\newtheorem{corollary}[theorem]{Corollary}

\newtheorem{problem}[theorem]{Problem}

\newenvironment{proof}[1][Proof]{\noindent\textbf{#1.} }{\ \rule{0.5em}{0.5em}}
\begin{document}

\author{Vadim E. Levit and David Tankus\\Department of Computer Science and Mathematics\\Ariel University, ISRAEL\\\{levitv, davidta\}@ariel.ac.il}
\date{}
\title{Well-dominated graphs without cycles of lengths $4$ and $5$}
\maketitle

\begin{abstract}
Let $G$ be a graph. A set $S$ of vertices in $G$ \textit{dominates} the graph
if every vertex of $G$ is either in $S$ or a neighbor of a vertex in $S$.
Finding a minimal cardinality set which dominates the graph is an
\textbf{NP}-complete problem. The graph $G$ is \textit{well-dominated} if all
its minimal dominating sets are of the same cardinality. The complexity status
of recognizing well-dominated graphs is not known. We show that recognizing
well-dominated graphs can be done polynomially for graphs without cycles of
lengths $4$ and $5$, by proving that a graph belonging to this family is
well-dominated if and only if it is well-covered.

Assume that a weight function $w$ is defined on the vertices of $G$. Then $G$
is $w$\textit{-well-dominated} if all its minimal dominating sets are of the
same weight. We prove that the set of weight functions $w$ such that $G$ is
$w$-well-dominated is a vector space, and denote that vector space by
$WWD(G)$. We prove that $WWD(G)$ is a subspace of $WCW(G)$, the vector space
of weight functions $w$ such that $G$ is $w$-well-covered. We provide a
polynomial characterization of $WWD(G)$ for the case that $G$ does not contain
cycles of lengths $4$, $5$, and $6$.

\end{abstract}


\section{Introduction}

\subsection{Definitions and Notations}

Throughout this paper $G$ is a simple (i.e., a finite, undirected, loopless
and without multiple edges) graph with vertex set $V(G)$ and edge set $E(G)$.

Cycles of $k$ vertices are denoted by $C_{k}$. When we say that $G$ does not
contain $C_{k}$ for some $k\geq3$, we mean that $G$ does not admit subgraphs
isomorphic to $C_{k}$. It is important to mention that these subgraphs are not
necessarily induced. Let $\mathcal{G}(\widehat{C_{i_{1}}},..,\widehat{C_{i_{k}%
}})$ be the family of all graphs which do not contain $C_{i_{1}},...,C_{i_{k}%
}$.

Let $u$ and $v$ be two vertices in $G$. The \textit{distance} between $u$ and
$v$, denoted $d(u,v)$, is the length of a shortest path between $u$ and $v$,
where the length of a path is the number of its edges. If $S$ is a non-empty
set of vertices, then the \textit{distance} between $u$ and $S$, denoted
$d(u,S)$, is defined by:
\[
d(u,S)=min\{d(u,s):s\in S\}.
\]
For every $i$, denote
\[
N_{i}(S)=\{x\in V(G):d(x,S)=i\},
\]
and
\[
N_{i}[S]=\{x\in V(G):d(x,S)\leq i\}.
\]
We abbreviate $N_{1}(S)$ and $N_{1}[S]$ to be $N(S)$ and $N[S]$, respectively.
If $S$ contains a single vertex, $v$, then we abbreviate $N_{i}(\{v\})$,
$N_{i}[\{v\}]$, $N(\{v\})$, and $N[\{v\}]$ to be $N_{i}(v)$, $N_{i}[v]$,
$N(v)$, and $N[v]$, respectively.

For every vertex $v \in V(G)$, the \textit{degree} of $v$ is $d(v) = |N(v)|$.
Let $L(G)$ be the set of all vertices $v \in V(G)$ such that either $d(v)=1$
or $v$ is on a triangle and $d(v)=2$.

\subsection{Well-Covered Graphs}

A set of vertices is \textit{independent} if its elements are pairwise
nonadjacent. Define $D(v) = N(v) \setminus N(N_{2}(v))$, and let $M(v)$ be a
maximal independent set of $D(v)$. An independent set of vertices is
\textit{maximal} if it is not a subset of another independent set. An
independent set is \textit{maximum} if $G$ does not admit an independent set
with a bigger cardinality. Denote $i(G)$ the minimal cardinality of a maximal
independent set in $G$, where $\alpha(G)$ is the cardinality of a maximum
independent set in $G$.

The graph $G$ is \textit{well-covered} if $i(G)=\alpha(G)$, i.e. all maximal
independent sets are of the same cardinality. The problem of finding a maximum
cardinality independent set $\alpha(G)$ in an input graph is \textbf{NP}%
-complete. However, if the input is restricted to well-covered graphs, then a
maximum cardinality independent set can be found polynomially using the
\textit{greedy algorithm}.

Let $w:V(G)\longrightarrow\mathbb{R}$ be a weight function defined on the
vertices of $G$. For every set $S\subseteq V(G)$, define $w(S)=\Sigma_{s\in
S}w(s)$. The graph $G$ is $w$\textit{-well-covered} if all maximal independent
sets are of the same weight. The set of weight functions $w$ for which $G$ is
$w$-well-covered is a \textit{vector space} \cite{cer:degree}. That vector
space is denoted $WCW(G)$ \cite{bnz:wcc4}.

Since recognizing well-covered graphs is \textbf{co-NP}-complete
\cite{cs:note} \cite{sknryn:compwc}, finding the vector space $WCW(G)$ of an
input graph $G$ is \textbf{co-NP}-hard. Finding $WCW(G)$ remains
\textbf{co-NP}-hard when the input is restricted to graphs with girth at least
$6$ \cite{lt:complexity}, and bipartite graphs \cite{lt:complexity}. However,
the problem is polynomially solvable for $K_{1,3}$-free graphs
\cite{lt:equimatchable}, and for graphs with a bounded maximal degree
\cite{lt:complexity}. For every graph $G$ without cycles of lengths $4$, $5$,
and $6$, the vector space $WCW(G)$ is characterized as follows.

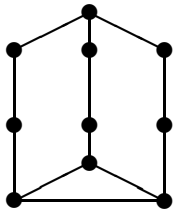
\begin{figure}[h]
\setlength{\unitlength}{1.0cm} \begin{picture}(3,3)\thicklines
\multiput(6,0.5)(0,1){3}{\circle*{0.2}}
\multiput(8,0.5)(0,1){3}{\circle*{0.2}}
\multiput(7,1.5)(0,1){2}{\circle*{0.2}}
\put(7,1){\circle*{0.2}}
\put(7,3){\circle*{0.2}}
\put(6,0.5){\line(0,1){2}}
\put(6,0.5){\line(1,0){2}}
\put(6,0.5){\line(2,1){1}}
\put(8,0.5){\line(0,1){2}}
\put(7,1){\line(0,1){2}}
\put(7,1){\line(2,-1){1}}
\put(6,2.5){\line(2,1){1}}
\put(7,3){\line(2,-1){1}}
\end{picture}\caption{The graph $T_{10}$.}%
\label{T10}%
\end{figure}

\begin{theorem}
\cite{lt:wc456} \label{wwc456} Let $G\in\mathcal{G}(\widehat{C_{4}%
},\widehat{C_{5}},\widehat{C_{6}})$ be a graph, and let $w:V(G)\longrightarrow
\mathbb{R}$. Then $G$ is w-well-covered if and only if one of the following holds:

\begin{enumerate}
\item $G$ is isomorphic to either $C_{7}$ or $T_{10}$ (see Figure \ref{T10}),
and there exists a constant $k \in\mathbb{R}$ such that $w \equiv k$.

\item The following conditions hold:

\begin{itemize}
\item $G$ is isomorphic to neither $C_{7}$ nor $T_{10}$.

\item For every two vertices, $l_{1}$ and $l_{2}$, in the same component of
$L(G)$ it holds that $w(l_{1})=w(l_{2})$.

\item For every $v \in V(G) \setminus L(G)$ it holds that $w(v) = w(M(v))$ for
some maximal independent set $M(v)$ of $D(v)$.
\end{itemize}
\end{enumerate}
\end{theorem}

Recognizing well-covered graphs is a restricted case of finding $WCW(G)$.
Therefore, for all families of graphs for which finding $WCW(G)$ is polynomial
solvable, recognizing well-covered graphs is polynomial solvable as well.
Recognizing well-covered graphs is \textbf{co-NP}-complete for $K_{1,4}$-free
graphs \cite{cst:structures}, but it is polynomially solvable for graphs
without cycles of lengths $3$ and $4$ \cite{fhn:wcg5}, for graphs without
cycles of lengths $4$ and $5$ \cite{fhn:wc45}, or for chordal graphs
\cite{ptv:chordal}.

\subsection{Well-Dominated Graphs}

Let $S$ and $T$ be two sets of vertices of the graph $G$. Then $S$
\textit{dominates} $T$ if $T\subseteq N[S]$. The set $S$ is
\textit{dominating} if it dominates all vertices of the graph. A dominating
set is \textit{minimal} if it does not contain another dominating set. A
dominating set is \textit{minimum} if $G$ does not admit a dominating set with
smaller cardinality. Let $\gamma(G)$ be the cardinality of a minimum
dominating set in $G$, and let $\Gamma(G)$ be the maximal cardinality of a
minimal dominating set of $G$. If $\gamma(G)=\Gamma(G)$ then the graph is
\textit{well-dominated}, i.e. all minimal dominating sets are of the same
cardinality. This concept was introduced in \cite{fhn:wd}, and further studied
in \cite{hhs:domination}. The fact that every maximal independent set is also
a minimal dominating set implies that
\[
\gamma(G)\leq i(G)\leq\alpha(G)\leq\Gamma(G)
\]
for every graph $G$. Hence, if $G$ is not well-covered, then it is not well-dominated.

\begin{theorem}
\cite{fhn:wd} \label{wdwc} Every well-dominated graph is well-covered.
\end{theorem}

In what follows, our main subject is the interplay between well-covered and
well-dominated graphs.

\begin{problem}
$WD$

\textit{Input:} A graph $G$.

\textit{Question:} Is $G$ well-dominated?
\end{problem}

It is even not known whether the $WD$ problem is in \textbf{NP}
\cite{d:networks}. However, the $WD$ problem is polynomial for graphs with
girth at least $6$ \cite{fhn:wd}, and for bipartite graphs \cite{fhn:wd}. We
prove that a graph without cycles of lengths $4$ and $5$ is well-dominated if
and only if it is well-covered. Consequently, by \cite{fhn:wc45}, recognizing
well-dominated graphs without cycles of lengths $4$ and $5$ is a polynomial task.

Let $w:V(G)\longrightarrow\mathbb{R}$ be a weight function defined on the
vertices of $G$. Then $G$ is $w$\textit{-well-dominated} if all minimal
dominating sets are of the same weight. Let $WWD(G)$ denote the set of weight
functions $w$ such that $G$ is $w$-well-dominated. It turns out that for every
graph $G$, $WWD(G)$ is a vector space. Moreover, if $G$ is $w$-well-dominated
then $G$ is $w$-well-covered, i.e., $WWD(G)$ $\subseteq WCW(G)$.

\begin{problem}
$WWD$

\textit{Input:} A graph $G$.

\textit{Output:} The vector space of weight functions $w$ such that $G$ is $w$-well-dominated.
\end{problem}

Finally, we supply a polynomial characterization of the $WWD$ problem, when
its input is restricted to $\mathcal{G}(\widehat{C_{4}},\widehat{C_{5}%
},\widehat{C_{6}})$.

\section{Well-Dominated Graphs Without $C_{4}$ and $C_{5}$}

A vertex $v$ is \textit{simplicial} if $N[v]$ is a clique. In \cite{fhn:wc45}
the family $F$ of graphs is defined as follows. A graph $G$ is in the family
$F$ if there exists $\{x_{1},...,x_{k}\}\subseteq V(G)$ such that $x_{i}$ is
simplicial for each $1\leq i\leq k$, and $\{N\left[  x_{i}\right]  :1\leq
i\leq k\}$ is a partition of $V(G)$. Well-covered graphs without cycles of
lengths $4$ and $5$ are characterized as follows.

\begin{theorem}
\label{wc456} \cite{fhn:wc45} Let $G\in\mathcal{G}(\widehat{C_{4}%
},\widehat{C_{5}})$ be a connected graph. Then $G$ is well-covered if and only
if one of the following holds:

\begin{enumerate}
\item $G$ is isomorphic to either $C_{7}$ or $T_{10}$.

\item $G$ is a member of the family $F$.
\end{enumerate}
\end{theorem}

Actually, under the restriction $G\in\mathcal{G}(\widehat{C_{4}}%
,\widehat{C_{5}})$, the families of well-covered and well-dominated graphs coincide.

\begin{theorem}
\label{gwd45} Let $G\in\mathcal{G}(\widehat{C_{4}},\widehat{C_{5}})$ be a
connected graph. Then $G$ is well-dominated if and only if it is well-covered.
\end{theorem}

\begin{proof}
By Theorem \ref{wdwc}, if $G$ is not well-covered then it is not well-dominated.

Assume $G$ is well-covered, and it should be proved that $G$ is
well-dominated. One can verify that $\gamma(C_{7}) = \Gamma(C_{7}) = 3$, and
$\gamma(T_{10}) = \Gamma(T_{10}) = 4$. Therefore, $C_{7}$ and $T_{10}$ are well-dominated.

By Theorem \ref{wc456}, it remains to prove that if $G$ is a member of $F$
then it is well-dominated. There exists $\{x_{1},...,x_{k}\}\subseteq V(G)$
such that $x_{i}$ is simplicial for each $1\leq i\leq k$, and $\{N[x_{i}%
]:1\leq i\leq k\}$ is a partition of $V(G)$. Define $V_{i}=N[x_{i}]$ for each
$1\leq i\leq k$. Let $S$ be a minimal dominating set of $G$. It is enough to
prove that $|S|=k$. The fact that $S$ dominates $x_{i}$ implies that $S\cap
V_{i}\neq\Phi$. Assume on the contrary that there exists $1\leq i\leq k$ such
that $\left\vert V_{i}\cap S\right\vert \geq2$. Let $S^{\prime}\subset S$ such
that $\left\vert S^{\prime}\cap V_{i}\right\vert =1$ for each $1\leq i\leq k$.
Clearly, $S^{\prime}$ dominates the whole graph, which is a contradiction.
Therefore, $\left\vert S\right\vert =k$, and $G$ is well-dominated.
\end{proof}

If $G\not \in \mathcal{G}(\widehat{C_{4}})$, then Theorem \ref{gwd45} does not
hold. Let $n\geq3$. Obviously, $K_{n,n}\in\mathcal{G}(\widehat{C_{5}})$, and
the cardinality of every maximal independent set of $K_{n,n}$ is $n$.
Therefore, $K_{n,n}$ is well-covered. However, there exists a minimal
dominating set of cardinality $2$. Therefore, $K_{n,n}$ is not well-dominated.

If $G\not \in \mathcal{G}(\widehat{C_{5}})$, then Theorem \ref{gwd45} does not
hold. Let $G$ be comprised of three disjoint $5$-cycles, $(x_{1}%
,...,x_{5}),(y_{1},...,y_{5}),(z_{1},...,z_{5})$, and a triangle $(x_{1}%
,y_{1},z_{1})$. Clearly, $G\in\mathcal{G}(\widehat{C_{4}})$, and every maximal
independent set contains $2$ vertices from each $5$-cycle. Hence, the
cardinality of every maximal independent set is $6$, and $G$ is well-covered.
However, $G$ is not well-dominated because it contains a minimal dominating
set of cardinality $7$: $\{x_{1},x_{2},x_{5},y_{3},y_{4},z_{3},z_{4}\}$.

\section{Weighted Well-Dominated Graphs}

\begin{theorem}
\label{wsgds} Let $G$ be a graph. Then the set of weight functions
$w:V(G)\longrightarrow\mathbb{R}$ such that $G$ is $w$-well-dominated is a
vector space.
\end{theorem}

\begin{proof}
Obviously, if $w_{0} \equiv0$ then $G$ is $w_{0}$-well-dominated.

Let $w_{1},w_{2}:V(G)\longrightarrow\mathbb{R}$, and assume that $G$ is
$w_{1}$-well-dominated and $w_{2}$-well-dominated. Then there exist two
constants, $t_{1}$ and $t_{2}$, such that $w_{1}(S)=t_{1}$ and $w_{2}%
(S)=t_{2}$ for every minimal dominating set $S$ of $G$. Let $\lambda
\in\mathbb{R}$, and let $w:V(G)\longrightarrow\mathbb{R}$ be defined by
$w(v)=w_{1}(v)+\lambda w_{2}(v)$ for every $v\in V(G)$. Then for every minimal
dominating set $S$ it holds that
\[
w(S)=%
{\displaystyle\sum\limits_{s\in S}}
w(s)=%
{\displaystyle\sum\limits_{s\in S}}
(w_{1}(s)+\lambda w_{2}(s))=%
{\displaystyle\sum\limits_{s\in S}}
w_{1}(s)+\lambda%
{\displaystyle\sum\limits_{s\in S}}
w_{2}(s)=t_{1}+\lambda t_{2},
\]
and $G$ is $w$-well-dominated.
\end{proof}

For every graph $G$, we denote the vector space of weight functions $w$ such
that $G$ is $w$-well-dominated by $WWD(G)$.

Let $G$ be a graph, and let $w:V(G)\longrightarrow\mathbb{R}$. Denote
$mDS_{w}(G)$, $MDS_{w}(G)$, $mIS_{w}(G)$, $MIS_{w}(G)$ the minimum weight of a
dominating set, the maximum weight of a minimal dominating set, the minimum
weight of a maximal independent set, and the maximum weight of an independent
set, respectively. 

The fact that every maximal independent set is also a minimal dominating set
implies that
\[
mDS_{w}(G)\leq mIS_{w}(G)\leq MIS_{w}(G)\leq MDS_{w}(G)
\]
for every graph $G$ and every weight function $w$ defined on its vertices.

If $mIS_{w}(G) = MIS_{w}(G)$ then $G$ is $w$-well-covered, and if $mDS_{w}(G)
= MDS_{w}(G)$ then $G$ is $w$-well-dominated. Theorem \ref{wdwc} is an
instance of the following.

\begin{corollary}
\label{wwdwwc} For every graph $G$ and for every weight function
$w:V(G)\longrightarrow\mathbb{R}$, if $G$ is $w$-well-dominated then $G$ is
$w$-well-covered, i.e., $WWD(G)$ is a subspace of $WCW(G)$.
\end{corollary}

\begin{figure}[h]
\setlength{\unitlength}{1.0cm} \begin{picture}(3,5.5)\thicklines
\put(2.25,5){\circle*{0.2}}
\put(2.25,5.3){\makebox(0,0){$x$}}
\put(2.25,5){\line(1,-2){0.75}}
\put(2.25,5){\line(-1,-2){0.75}}
\put(6.75,5){\circle*{0.2}}
\put(6.75,5.3){\makebox(0,0){$y_{1}$}}
\put(6.75,5){\line(1,-2){0.75}}
\put(6.75,5){\line(-1,-2){0.75}}
\multiput(1.5,3.5)(1.5,0){8}{\circle*{0.2}}
\put(12,3.8){\makebox(0,0){$y_{4}$}}
\put(1.5,3.5){\line(1,0){10.5}}
\multiput(0.75,2)(0.75,0){5}{\circle*{0.2}}
\put(0.75,1.6){\makebox(0,0){$y_{5}$}}
\put(3,1.6){\makebox(0,0){$y_{6}$}}
\put(3.75,1.6){\makebox(0,0){$y_{7}$}}
\put(1.5,3.5){\line(-1,-2){0.75}}
\put(1.5,3.5){\line(0,-1){1.5}}
\put(1.5,3.5){\line(1,-2){0.75}}
\put(3,3.5){\line(1,-2){0.75}}
\put(3,3.5){\line(0,-1){1.5}}
\multiput(5.25,2)(0.75,0){4}{\circle*{0.2}}
\put(6,3.5){\line(-1,-2){0.75}}
\put(6,3.5){\line(0,-1){1.5}}
\put(7.5,3.5){\line(-1,-2){0.75}}
\put(7.5,3.5){\line(0,-1){1.5}}
\put(1.5,0.5){\circle*{0.2}}
\put(1.5,0.2){\makebox(0,0){$y_{8}$}}
\put(2.25,0.5){\circle*{0.2}}
\put(1.5,0.5){\line(0,1){1.5}}
\put(2.25,0.5){\line(0,1){1.5}}
\put(2.25,0.5){\line(1,0){3}}
\multiput(5.25,0.5)(0.75,0){4}{\circle*{0.2}}
\put(6,0.2){\makebox(0,0){$y_{9}$}}
\put(6.75,0.2){\makebox(0,0){$y_{10}$}}
\put(7.5,0.2){\makebox(0,0){$y_{11}$}}
\multiput(5.25,0.5)(0.75,0){4}{\line(0,1){1.5}}
\put(10.5,3.5){\line(-1,2){0.75}}
\put(10.5,3.5){\line(1,2){0.75}}
\put(9.75,5){\circle*{0.2}}
\put(11.25,5){\circle*{0.2}}
\put(9.75,5.3){\makebox(0,0){$y_{2}$}}
\put(11.25,5.3){\makebox(0,0){$y_{3}$}}
\put(9.75,5){\line(1,0){1.5}}
\end{picture}\caption{An example of the definition of $L^{\ast}(G)$. In this
graph, $G$, it holds that $L^{\ast}(G)=\{y_{1},...,y_{11}\}$ and
$L(G)\setminus L^{\ast}(G)=\{x\}$. Let $w:V(G)\longrightarrow\mathbb{R}$. By
Theorem \ref{wwc456}, $G$ is $w$-well-covered if and only if $w(y_{2}%
)=w(y_{3})$ and $w(v)=w(M(v))$ for every $v\in V(G)\setminus L(G)$. By Theorem
\ref{wwd456}, $G$ is $w$-well-dominated if and only if $G$ is $w$-well-covered
and $w(x)=0$.}%
\label{Lstar}%
\end{figure}
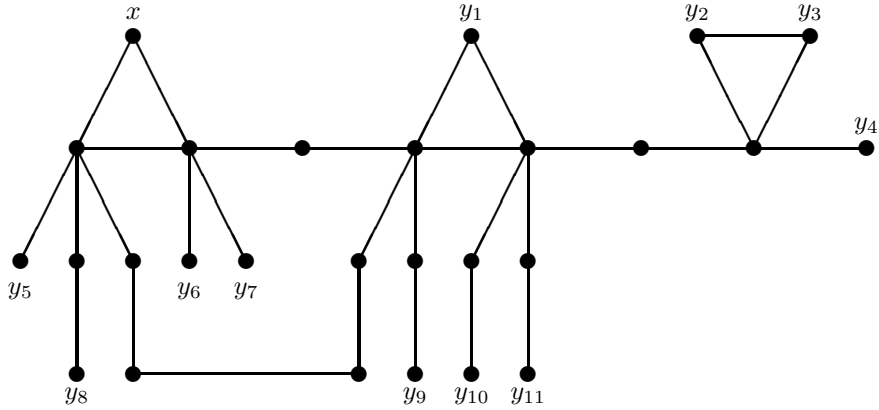

Let $L^{\ast}(G)$ be the set of all vertices $v\in V(G)$ such that either

\begin{itemize}
\item $d(v)=1$;

or

\item the following conditions hold:

\begin{itemize}
\item $d(v)=2$;

\item $v$ is on a triangle, $(v,v_{1},v_{2})$;

\item Every maximal independent set of $V(G)\setminus N_{2}[v]$ dominates at
least one of $N(v_{1})\cap N_{2}(v)$ and $N(v_{2})\cap N_{2}(v)$.
\end{itemize}
\end{itemize}

Note that $L^{\ast}(G)\subseteq L(G)$ (see Figure \ref{Lstar}). Moreover,
$v\in L(G)\setminus L^{\ast}(G)$ if and only if the following conditions hold:

\begin{itemize}
\item $d(v)=2$

\item $v$ is on a triangle, $(v,v_{1},v_{2})$.

\item There exists a maximal independent set of $V(G)\setminus N_{2}[v]$ which
dominates neither $N(v_{1})\cap N_{2}(v)$ nor $N(v_{2})\cap N_{2}(v)$.
\end{itemize}

\begin{theorem}
\label{wwd456} Let $G\in\mathcal{G}(\widehat{C_{4}},\widehat{C_{5}%
},\widehat{C_{6}})$ be a connected graph, and let $w:V(G)\longrightarrow
\mathbb{R}$. Then $G$ is $w$-well-dominated if and only if one of the
following holds:

\begin{enumerate}
\item $G$ is isomorphic to either $C_{7}$ or $T_{10}$ (see Figure \ref{T10}),
and there exists a constant $k \in\mathbb{R}$ such that $w \equiv k$.

\item The following conditions hold:

\begin{enumerate}
\item $G$ is isomorphic to neither $C_{7}$ nor $T_{10}$.

\item For every two vertices, $l_{1}$ and $l_{2}$, in the same component of
$L(G)$ it holds that $w(l_{1})=w(l_{2})$.

\item $w(v)=0$ for every vertex $v \in L(G) \setminus L^{*}(G)$.

\item For every $v \in V(G) \setminus L(G)$ it holds that $w(v) = w(M(v))$ for
some maximal independent set $M(v)$ of $D(v)$.
\end{enumerate}
\end{enumerate}
\end{theorem}

\begin{proof}
The following cases are considered.

\textsc{Case 1: $G$ is isomorphic to either $C_{7}$ or $T_{10}$.} If there
does not exist a constant $k\in\mathbb{R}$ such that $w\equiv k$, then by
Theorem \ref{wwc456}, $G$ is not $w$-well-covered. By Corollary \ref{wwdwwc},
$G$ is not $w$-well-dominated. Suppose that $w\equiv k$ for some
$k\in\mathbb{R}$. If $G$ is isomorphic to $C_{7}$, then the cardinality of
every minimal dominating set is $3$. Hence, $mDS_{w}(C_{7})=MDS_{w}(C_{7}%
)=3k$, and $G$ is $w$-well-dominated. If $G$ is isomorphic to $T_{10}$, then
the cardinality of every minimal dominating set is $4$. Hence, $mDS_{w}%
(T_{10})=MDS_{w}(T_{10})=4k$, and $G$ is $w$-well-dominated.

\textsc{Case 2: $L(G)=V(G)$.} In this case $G$ is a complete graph with at
most $3$ vertices. In that case the cardinality of every minimal dominating
set is 1. Therefore, $G$ is $w$-well-dominated if and only if there exists a
constant $k\in\mathbb{R}$ such that $w\equiv k$. In this case $mDS_{w}%
(G)=MDS_{w}(G)=k$.

\textsc{Case 3: $L(G)\neq V(G)$ and $G$ is isomorphic to neither $C_{7}$ nor
$T_{10}$.} Let $N(L^{\ast}(G))=\{v_{1},...,v_{k}\}$. Then
\[
L^{\ast}(G)\subseteq%
{\displaystyle\bigcup\limits_{1\leq i\leq k}}
D(v_{i})\subseteq L(G).
\]
For each $1\leq i\leq k$ let $V_{i}=\{v_{i}\}\cup D(v_{i})$.

Assume that Condition 2 holds. Let $S$ be a minimal dominating set of $G$.
Then $S\cap V_{i}$ is either $v_{i}$ or $M(v_{i})$. Hence, $w(S\cap
V_{i})=w(v_{i})$ for every $1\leq i\leq k$. Let $1\leq i<j\leq k$. Then
$V_{i}\cap V_{j}\subseteq L(G)\setminus L^{\ast}(G)$. Hence, $w(x)=0$ for each
$x\in V_{i}\cap V_{j}$. The fact that $w(v)=0$ for every $v\in V\setminus
N[L(G)]$ implies that
\begin{gather*}
w(S)=w(S\setminus(%
{\displaystyle\bigcup\limits_{1\leq i\leq k}}
V_{i}))+%
{\displaystyle\sum\limits_{1\leq i\leq k}}
w(S\cap V_{i})-%
{\displaystyle\sum\limits_{1\leq i\leq j\leq k}}
w(S\cap V_{i}\cap V_{j})=\\
=0+%
{\displaystyle\sum\limits_{1\leq i\leq k}}
w(v_{i})-0=%
{\displaystyle\sum\limits_{1\leq i\leq k}}
w(v_{i}).
\end{gather*}
Hence,
\[
mDS_{w}(G)=MDS_{w}(G)=%
{\displaystyle\sum\limits_{1\leq i\leq k}}
w(v_{i}),
\]
and $G$ is $w$-well-dominated.

Assume that $G$ is $w$-well-dominated. Then, by Corollary \ref{wwdwwc}, $G$ is
$w$-well-covered. By Theorem \ref{wwc456}, Conditions 2a, 2b and 2d hold. It
remains to prove that Condition 2c holds as well. Let $v\in L(G)\setminus
L^{\ast}(G)$. We prove that $w(v)=0$. Let $N(v)=\{v_{1},v_{2}\}$, and let $S$
be a maximal independent set of $G\setminus N_{2}[v]$ which dominates neither
$N(v_{1})\cap N_{2}(v)$ nor $N(v_{2})\cap N_{2}(v)$. For each $1\leq i\leq2$
let $S_{i}$ be a maximal independent set of $(N(v_{i})\cap N_{2}(v))\setminus
N(S)$. Define $T_{i}=S\cup S_{2-i}\cup\{v_{i}\}$ for $i=1,2$. Define
$T_{3}=S\cup S_{1}\cup S_{2}\cup\{v\}$ and $T_{4}=S\cup\{v_{1},v_{2}\}$.
Clearly, $T_{1}$, $T_{2}$, $T_{3}$ and $T_{4}$ are minimal dominating sets of
$G$.

For each $i=1,2$ the fact that $w(T_{i})=w(T_{3})$ implies $w(S_{i}%
\cup\{v\})=w(v_{i})$. Therefore $w(S_{i})+w(v)=w(v_{i})$. The fact that
$w(T_{3})=w(T_{4})$ implies $w(S_{1}\cup S_{2}\cup\{v\})=w(\{v_{1},v_{2}\})$.
Therefore, $w(S_{1})+w(S_{2})+w(v)=w(v_{1})+w(v_{2})$. Hence, $w(S_{1}%
)+w(S_{2})+w(v)=w(S_{1})+w(S_{2})+2w(v)$. Thus $w(v)=0$.
\end{proof}

\begin{corollary}
\label{dimwwd} $dim(WWD(G))=|L^{\ast}(G)|$ for every graph $G\in
\mathcal{G}(\widehat{C_{4}},\widehat{C_{5}},\widehat{C_{6}})$ .
\end{corollary}

\begin{corollary}
\label{wwdeqwwc} Suppose $G\in\mathcal{G}(\widehat{C_{4}},\widehat{C_{5}%
},\widehat{C_{6}})$. If $L^{\ast}(G)=L(G)$, then $WWD(G)=WCW(G)$. Otherwise,
$WWD(G)\subsetneqq WCW(G)$.
\end{corollary}

Combining Corollaries \ref{dimwwd}, \ref{wwdeqwwc} with Algortihm 20 from
\cite{lt:wc456} we obtain the following.

\begin{corollary}
If $G\in\mathcal{G}(\widehat{C_{4}},\widehat{C_{5}},\widehat{C_{6}})$, then
\[
\left\vert L^{\ast}(G)\right\vert =dim(WWD(G))\leq dim(WCW(G))=\alpha\left(
G\left[  L(G)\right]  \right)  .
\]

\end{corollary}

Theorem \ref{wwd456} does not hold if $G\not \in \mathcal{G}(\widehat{C_{6}}%
)$. Let $G$ be the graph with two edge disjoint $6$-cycles, $(v_{1}%
,...,v_{6})$ and $(v_{6},...,v_{11})$. Clearly, $G\in\mathcal{G}%
(\widehat{C_{4}},\widehat{C_{5}})$ and $L(G)=L^{\ast}(G)=\Phi$. However, the
vector space $WWD(G)$ is the set of all functions $w:V(G)\longrightarrow
\mathbb{R}$ which satisfy

\begin{enumerate}
\item $w(v_{1}) = w(v_{2}) = -w(v_{4}) = -w(v_{5})$

\item $w(v_{7}) = w(v_{8}) = -w(v_{10}) = -w(v_{11})$

\item $w(v_{3}) = w(v_{6}) = w(v_{9}) = 0$
\end{enumerate}

\section{Future Work}

The main findings of the paper stimulate us to discover more cases, where the
$WD$ and/or $WWD$ problems can be solved polynomially.

We have proved that if $G\in\mathcal{G}(\widehat{C_{4}},\widehat{C_{5}})$,
then $G$ is well-dominated if and only if it is well-covered. It motivates the following.

\begin{problem}
Characterize all graphs, which are both well-covered and well-dominated.
\end{problem}

We have also shown that if $G\in\mathcal{G}(\widehat{C_{4}},\widehat{C_{5}%
},\widehat{C_{6}})$ and $L^{\ast}(G)=L(G)$, then $WCW(G)=WWD(G)$. Thus one may
be interested in approaching the following.

\begin{problem}
Characterize all graphs, where the equality $WCW(G)=WWD(G)$ holds.
\end{problem}

\end{document}